\documentstyle[12pt]{article}
\newcommand{\bmu}{{\wb \mu}}
\newcommand{\bnu}{{\wb \nu}}
\newcommand{\wb}{\bar}
\def\vev#1{\langle #1 \rangle}

\newcommand{\be}{\begin{equation}}
\newcommand{\ee}{\end{equation}}
\newcommand{\wt}{\widetilde}
\newcommand{\ben}{\begin{eqnarray}\displaystyle}
\newcommand{\een}{\end{eqnarray}}
\newcommand{\refb}[1]{(\ref{#1})}

\begin{document}

{}~ \hfill\vbox{\hbox{hep-th/9511026}%\hbox{MRI-PHY/27/95}
}\break

\vskip 3.5cm

\centerline{\large \bf U-DUALITY AND INTERSECTING D-BRANES}

\vspace*{6.0ex}

\centerline{\large \rm Ashoke Sen\footnote{On leave of absence from Tata
Institute of Fundamental Research, Homi Bhabha Road, Bombay 400005, INDIA}
\footnote{E-mail: sen@mri.ernet.in, sen@theory.tifr.res.in}}

\vspace*{1.5ex}

\centerline{\large \it Mehta Research Institute of Mathematics}
 \centerline{\large \it and Mathematical Physics}

\centerline{\large \it 10 Kasturba Gandhi Marg, Allahabad 211002, INDIA}

\vspace*{4.5ex}

\centerline {\bf Abstract}

Spectrum of elementary string states in type II string theory compactified
on a torus contains short multiplets which are invariant under only one
quarter of the space-time supersymmetry generators.  $U$-duality
transformation converts these states into bound states of Dirichlet branes
which wrap around intersecting cycles of the internal torus. We study a
class of these bound states that are dual to the elementary string states
at the first excited level, and argue that the degeneracy of these bound
states is in agreement with the $U$-duality prediction.

\vfill \eject

The spectrum of type IIA or IIB string theory compactified on a torus
contains single particle states in the ultra-short (256 dimensional)
multiplet. These states carry equal amount of left and right moving
charges and are in the ground state of {\it both}, the left and the right
moving oscillators. As a result they always have unit multiplicity. There
are also states in the short multiplet (of dimension $(16)^3$) which can
carry different amounts of left and right moving charges, and are in the
ground state of {\it either} the left {\it or} the right moving
oscillators. For definiteness, we shall assume from now on that the right
moving oscillators are in their ground state. The degeneracy of these
states depends on the level of the left moving oscillator state. If $\vec
Q_R$ and $\vec Q_L$ denote the right and the left moving charge vectors,
and $N_L$ denotes the level of the left moving part of the state, then,
with suitable normalization, the mass of the state (in the Ramond-Ramond
(RR) sector) is given by:
\be \label{e1}
M^2 = {\vec Q_R^2\over 2} = {\vec Q_L^2\over 2} +N_L \, .
\ee
Similar formula exits in the Neveu-Schwarz (NS) sector, but we do not need
to write it down. We shall take eq.\refb{e1} to be the defining equation
for $N_L$ for given $\vec Q_L$, $\vec Q_R$. The total degeneracy $d(N_L)$
of such states for a given values of $\vec Q_L$ and $\vec Q_R$, can be
computed from the formula:
\be \label{e2}
\sum_{N_L=0}^\infty d(N_L) q^{N_L} = 256 \times \prod_{n=1}^\infty
\bigg({1+q^n \over 1-q^n} \bigg)^8\, .
\ee
In order to find the number of short multiplets, one needs to divide the
value of $d(N_L)$ obtained this way by $(16)^3$, which is the dimension of
the supermultiplet. The degeneracies of some of the low lying states are
$d(1)=(16)^3$, $d(2)= 9\times (16)^3$ etc.

The charges $\vec Q_R$ and $\vec Q_L$ referred to above couple to the
$U(1)$ gauge fields arising in the NS-NS sector of the theory.  Since
$U$-duality\cite{HULL} transforms the states carrying NS-NS charges to
states carrying RR charges in general, and since the latter states have
been shown to arise from Dirichlet branes (D-branes)\cite{POLNEW} wrapped
around various internal cycles, we expect that we should be able to
reproduce the values of $d(N_L)$ quoted above by working out the
degeneracies of the states of $D$-branes. This analysis has been made
possible by the recent discovery of Witten\cite{WITTENNEW} that the
dynamics of collective coordinates of $n$ parallel $D$-branes is described
by a supersymmetric $U(n)$ gauge theory.\footnote{Another approach to
analyzing bound states of D-brane states has been given in \cite{LI}.}
Using this collective coordinate description, we have argued in a previous
paper\cite{DBRANE} that for ultra-short multiplets, the degeneracy of
D-brane states agrees with the prediction of $U$-duality.  In a recent
paper\cite{BERVAFA}  Bershadsky, Sadov, and Vafa have generalized Witten's
result to the case of intersecting $D$-branes. In this paper we shall use
this result to analyze the degeneracy of short multiplets.  In
particular, we shall work out the degeneracy of D-brane states dual to the
$N_L=1$ states and argue that the number is indeed equal to $d(1)$, in
agreement with the prediction of $U$-duality.

In order to be more specific, let us consider type IIA string theory
compactified on a four torus $T^4$, with each of the four internal circles
having self-dual radius. We shall denote the compact directions by $x^m$
for ($6\le m\le 9$), and the non-compact directions by $x^\bmu$ for ($0\le
\bmu\le 5$). The index $\mu$ will run over all values. This 6-dimensional
theory has eight $U(1)$ gauge fields coming from the NS-NS sector, and
eight $U(1)$ gauge fields coming from the RR sector. The $U$-duality group
of this theory has a specific $Z_2$ element, which changes all the NS-NS
gauge fields to RR gauge fields and vice versa, and acts as a triality
rotation on the $T$-duality group $SO(4,4;Z)$\cite{SENVAFA}. We shall use
this $Z_2$ element to convert a state carrying purely NS-NS charge to a
state carrying purely RR charge.

In order to be more specific about the action of this $Z_2$ element on
various charges, we need to choose a basis for the charge vector.  A
suitable choice of basis will be as follows. If $p_i$ and $w_i$ denote the
momentum and winding number associated with the internal directions $x^i$,
then we represent the charge vector as
\be \label{e2a}
Q=\pmatrix{p_6 \cr w_6 \cr \cdot \cr \cdot \cr p_9\cr w_9\cr}\, .
\ee
In this basis the inner product of two charge vectors is computed from the
metric,
\be \label{e3}
L=\pmatrix{\sigma_1 &&& \cr & \sigma_1 && \cr && \sigma_1 & \cr &&&
\sigma_1 \cr}\, , \qquad \sigma_1 = \pmatrix{ 0 & 1 \cr 1 & 0}\, .
\ee
In particular, $\vec Q_L$ and $\vec Q_R$ will denote the projection of the
charge vector to the subspace with $L$ eigenvalues $-1$ and $+1$
respectively. Also,
\be \label{e4}
\vec Q_R^2 - \vec Q_L^2 = Q^T L Q\, .
\ee
Consider now the charge vector:
\be \label{e5}
\pmatrix{m \cr n\cr 0 \cr \cdot \cr \cdot \cr 0}\, .
\ee
For this $Q^TLQ=2mn$. Using eq.\refb{e1} we see that for a short
multiplet, a state carrying this charge vector has $N_L=Q^TLQ/2=mn$. One
can compute the degeneracy of such states in the elementary string
spectrum from eq.\refb{e2}. Also the mass of this state (which we shall
refer to as $(m,n)$ state) is given by:
\be \label{e5a}
M(m,n) = {1\over \sqrt 2} |\vec Q_R| = {1\over 2} (|m|+|n|)\, .
\ee
This shows that these states are only marginally stable. In particular, an
$(m,n)$ state can decay into an $(m,0)$ state and a $(0,n)$ state at rest.
We shall come back to this point later.

What will be the dual of these states under the $Z_2$ transformation
mentioned above? The gauge fields in the RR sector come from the
components $C_{mn\bmu}$ of the rank three anti-symmetric tensor fields in
the IIA theory, as well as the gauge field $A_\bmu$ and the dual $\wt
C_\bmu$ of $C_{\bmu\bnu\bar\rho}$. In this basis, the inner product matrix
for the RR charges is given as follows. It pairs $A_\bmu$ charge with $\wt
C_{\bmu}$ charge. For $C_{mn\bmu}$ charge the inner product matrix is
simply the intersection matrix of the corresponding two cycles on the
torus. The $Z_2$ transformation mapping the RR gauge fields to NS-NS gauge
fields must preserve the inner product matrix. Thus we can choose the
$Z_2$ to induce the following map between the gauge fields (up to sign)
\ben \label{e6}
C_{89\bmu} \leftrightarrow G_{6\bmu} & \qquad &
C_{67\bmu} \leftrightarrow B_{6\bmu} \cr
C_{78\bmu} \leftrightarrow G_{7\bmu} & \qquad &
C_{96\bmu} \leftrightarrow B_{7\bmu} \cr
C_{68\bmu} \leftrightarrow G_{8\bmu} & \qquad &
C_{79\bmu} \leftrightarrow B_{8\bmu} \cr
A_{\bmu} \leftrightarrow G_{9\bmu} & \qquad &
\wt C_{\bmu} \leftrightarrow B_{9\bmu} \, .
\een
In particular, the momentum $p^6$ that couples to $G_{6\bmu}$ will be
mapped to $C_{89\bmu}$ charge and the winding $w^6$ that couples to
$B_{6\bmu}$ will be mapped to $C_{67\bmu}$ charge. Thus the $(m,n)$ state
will be mapped to a state with $m$ units of $C_{89\bmu}$ charge and $n$
units of $C_{67\bmu}$ charge. As shown in ref.\cite{POLNEW}, such a state
can arise from a state with $(m+n)$ Dirichlet membranes, with $m$ of them
wrapping around the 8-9 cycle of the torus, and $n$ of them wrapping
around the 6-7 cycle of the torus. In order to prove the invariance of the
spectrum under $U$-duality, one needs to show that this system contains
supersymmetric bound states with degeneracy $d(mn)$.

We shall analyze the $(1,1)$ case in detail and show that the degeneracy
of states of the D-membranes agree with $d(1)$ computed from elementary
string spectrum. But before we proceed we need to resolve the usual
problem with marginally stable states. Since a $(1,1)$ state can decay
into a $(1,0)$ state and a $(0,1)$ state at rest, there is no energy
barrier against pulling the two D-membranes away from each other. This
makes the analysis difficult.\footnote{Although in principle one can
directly study the question of existence of these marginally stable bound
states by mapping this problem to a supersymmetric quantum mechanics
problem, it has been pointed out by C. Vafa that naive computation of
Witten index in these models might not always agree with the predictions
of $U$-duality\cite{VAFAP}. This could be related to subtleties that might
be present at large separation. For this purpose we shall find it much
more convenient to work with the absolutely stable bound states described
below.} To get around this problem, we use the trick used in
ref.\cite{DBRANE}. Namely, we compactify one more direction (say $x^1$) so
that the momentum along $x^1$ is quantized, and look at the sector
carrying odd units of momentum in this direction. If a marginally stable
bound state of two D- membranes of the type discussed above exist in the 6
dimensional theory, then the Kaluza-Klein modes of this state, carrying
odd units of momentum along the $x^1$ direction, will be absolutely stable
in the resulting 5 dimensional theory. We shall look for these absolutely
stable states in the spectrum.

As in ref.\cite{DBRANE} we shall simplify the analysis by performing a
$T$-duality transformation involving the coordinate $x^1$. Let us denote
the coordinates in this new theory by $y^\mu$. This duality transformation
will convert the momentum along $x^1$ into winding
number\footnote{Throughout this paper, winding number will refer to the
winding number of an elementary string that couples to the $B_{\mu\nu}$
field.} along $y^1$, and convert a membrane wrapped around $m$-$n$ plane
into a three brane wrapped around the $m$-$n$-1 plane. Thus the problem
that we have is that of a three brane $D_1$ wrapped around the 8-9-1 plane
interacting with a three brane $D_2$ wrapped around the 6-7-1 plane. We
want to look for supersymmetric ground states of this system carrying $k$
units of winding along the $y^1$ direction, where $k$ is odd. The dynamics
of collective coordinates of this system has been given in
ref.\cite{BERVAFA}. Since the configuration described above has
translation invariance only along the $y^0$ and $y^1$ direction, it is
described by a $(1+1)$ dimensional field theory with base space labelled
by the coordinates $y^0$ and $y^1$. The theory is an $N=4$ supersymmetric
gauge theory (which can be regarded as the dimensional reduction of an
$N=2$ theory in four dimensions) with an $U(1)_1\times U(1)_2$ gauge
group. Besides the $U(1)$ vector multiplets which we shall denote by $A_1$
and $A_2$, the theory contains a pair of gauge neutral hypermultiplets
$\Phi_1$ and $\Phi_2$, and a hypermultiplet $Q$ of charge $(+1,-1)$ under
the $U(1)_1\times U(1)_2$ gauge group. For our purpose it will be
convenient to work with the diagonal sum and difference of the two $U(1)$
groups which we shall denote by $U(1)_c$ and $U(1)_r$. The corresponding
vector multiplets will be denoted by $A_c$ and $A_r$. (Here $c$ stands for
center of mass and $r$ for relative coordinates, for reasons that will be
explained soon). Then the gauge group is $\big(U(1)_c\times
U(1)_r\big)/Z_2$, and the hypermultiplet $Q$ has charge $(0,2)$ under
$U(1)_c\times U(1)_r$. Of these set of fields, the gauge multiplet $A_c$
and the hyper-multiplets $\Phi_1$ and $\Phi_2$ have no interactions,  the
only interaction of the theory comes from the standard gauge coupling
between the vector multiplet $A_r$ and the charged hypermultiplet $Q$.

Before we proceed further, let us give a physical interpretation of the
various bosonic fields, at least in the region where the two three branes
$D_1$ and $D_2$ are well separated in the physical space. We start with
the neutral hypermultiplets. Of the four scalars in the hypermultiplet
$\Phi_1$, two denote the $y^6$ and $y^7$ coordinates of the three brane
$D_1$ lying in the 8-9-1 plane. The other two are coordinates conjugate to
the winding number along the $8$ and the $9$ direction. Similarly, of the
four scalars in the hypermultiplet $\Phi_2$, two correspond to the $y^8$
and $y^9$ coordinates of the three brane $D_2$ lying in the 6-7-1 plane,
and two are conjugate to the winding number carried by $D_2$ along $y^6$
and $y^7$.  Since the winding as well as momenta along the $y^6,\ldots
y^9$ directions are quantized, we see that we should interprete the scalar
fields in both the hypermultiplets $\Phi_1$ and $\Phi_2$ as compact
coordinates in the field space.

A vector multiplet in two dimensions has four scalars and one vector
field. The four scalars in the $U(1)_1$ vector multiplet correspond to the
coordinates $(y^2,y^3,y^4,y^5)$ of the three brane $D_1$. The total
electric flux associated with the $U(1)_1$ gauge field along $y^1$ may be
interpreted as the winding number carried by $D_1$ along $y^1$, $-$ this
situation is identical to the one described in ref.\cite{WITTENNEW}.
Similarly the four scalars in the $U(1)_2$ vector multiplet correspond to
the coordinates $(y^2,y^3,y^4,y^5)$ of the three brane $D_2$, and the
total electric flux associated with the $U(1)_2$ gauge field along $y^1$
may be interpreted as the winding number carried by $D_2$ along $y^1$. In
terms of the vector multiplets $A_c$ and $A_r$ this means that the scalar
fields in $A_c$ denote the center of mass coordinates of $D_1$ and $D_2$
along $y^2, \ldots y^5$, the scalar fields in $A_r$ denote the relative
coordinates along $y^2, \ldots y^5$, the $U(1)_c$ electric flux along
$y^1$ denotes the total winding number of the system along $y^1$, and the
electric flux of $U(1)_r$ along $y^1$ measures the difference between the
winding numbers carried by the two Dirichlet three branes.

Finally we turn to the hypermultiplet $Q$. These fields do not have a
simple interpretation as space-time coordinates of the three branes, as
can be seen from the fact that due to their coupling to $A_r$, they become
very heavy when the relative separation between $D_1$ and $D_2$ is large.
These correspond to open string states that start on $D_1$ and end on
$D_2$ or vice versa.\footnote{Physically, the absence of coupling between
the neutral hypermultiplets $\Phi_1$, $\Phi_2$ and the charged
hypermultiplet $Q$ may be understood as follows. Mass of $Q$ should vanish
whenever the two D-branes intersect. This happens whenever the spatial
separation between the two D-branes along $y^2,\ldots y^5$ directions
vanish. Since these spatial separations are represented by scalar
components of the vector multiplet $A_r$, the mass of $Q$ cannot depend on
the vacuum expectation values of the hypermultiplets $\Phi_1$ and
$\Phi_2$.}

Let us now turn to the quantization of this system. First we discuss the
free part of the theory, the one involving $A_c$, $\Phi_1$ and $\Phi_2$.
Since we are looking for states that do not carry any momentum or winding
along any of the directions $y^6,\ldots y^9$, we can take the momenta
conjugate to the scalar components of $\Phi_1$ and $\Phi_2$ to be zero.
This gives a unique state from this sector. Similarly if we are
considering the system at rest we can take the total momenta, which are
conjugate to the four scalar components of $A_c$, to vanish. Finally we
can take the system to be in the eigenstate of the electric flux of
$U(1)_c$ gauge field with eigenvalue $k$, where $k$ is the required
winding number along $y^1$. For reasons that has been explained before, we
shall take $k$ to be odd. Thus the quantization of the bosonic part of the
free system gives a unique state for every value of $k$.

What about the fermionic part? Each neutral hypermultiplet contains eight
real fermionic coordinates, and the $U(1)_c$ vector multiplet also
contains  eight real fermionic coordinates. This gives rise to 24
fermionic zero modes in total. Quantization of this gives rise to a
$2^{12}= (16)^3$ fold degeneracy.

Thus if $K$ denotes the number of supersymmetric ground states of the
interacting system, involving the vector multiplet $A_r$ and the charged
hypermultiplet $Q$, then the total degeneracy of this state will be given
by:
\be \label{e7}
(16)^3 \cdot K\, .
\ee
This number then needs to be compared with $d(1)$. We have already stated
that $d(1)=(16)^3$; thus we see that the prediction of $U$-duality is
\be \label{e8}
K=1\, .
\ee
We shall now proceed to verify this prediction. The approach that we shall
use is identical to the one taken in ref.\cite{WITTENNEW}. The interacting
system can be viewed as the dimensional reduction of a four dimensional
theory, containing an $N=2$ $U(1)$ vector supermultiplet interacting with
an $N=2$ hypermultiplet of $U(1)$ charge two. We are looking for a
supersymmetric ground state in this theory in the sector that carries odd
unit of $U(1)_r$ flux along $y^1$; this requirement comes from the fact
that the gauge group is really $\big(U(1)_c\times U(1)_r\big)/Z_2$, and
hence the presence of an odd unit of $U(1)_c$ flux forces us to have an
odd unit of $U(1)_r$ flux as well. Viewed as an $N=1$ supersymmetric
theory, this corresponds to an $N=1$ supersymmetric $U(1)$ gauge theory,
with a $U(1)$ neutral chiral superfield $\Phi$, and a pair of chiral
superfields $\Lambda$, $\wb\Lambda$ with $U(1)$ charge $\pm 2$,
interacting through the superpotential:
\be \label{e9}
W_0=\Phi\Lambda\wb\Lambda\, .
\ee
The classical theory has a flat direction, since $\vev{\Phi}$ is
undetermined. However, for large $\vev{\Phi}$ the fields $\Lambda$,
$\wb\Lambda$ become heavy, and the resulting low energy theory is that of
a supersymmetric $U(1)$ gauge theory with no matter. The $U(1)_r$ electric
flux along the $y^1$ direction is not screened, and as a result this field
configuration has non-zero energy and breaks supersymmetry.  This shows
that there is a finite energy barrier against taking $\Phi$ to be large.
As in ref.\cite{WITTENNEW}, we shall assume that due to the presence of
the energy barrier, we can add mass terms for various fields in the
superpotential without changing the number of supersymmetric ground
states. Thus we work with the modified superpotential:
\be \label{e10}
W = \Phi \Lambda \wb \Lambda + {1\over 2} \epsilon \Phi^2 + \eta \Lambda
\wb\Lambda\, .
\ee
Possible supersymmetric ground states of this system corresponding to the
critical points of $W$ are obtained by solving the equations:
\be \label{e11}
\epsilon \Phi + \wb\Lambda \Lambda = 0\, , \qquad
(\Phi+\eta)\Lambda=0=(\Phi+\eta) \wb\Lambda\, .
\ee
For the trivial solution, given by $\Phi=\Lambda=\wb\Lambda=0$, the fields
$\Phi$, $\Lambda$ and $\wb\Lambda$ are all massive, and $U(1)$ is
unbroken. Thus the background electric field remains unscreened, costing a
finite energy, and hence supersymmetry is broken in this vacuum. The
non-trivial solution, up to (complexified) $U(1)$ gauge transformation, is
given by
\be \label{e12}
\Phi=-\eta, \qquad \Lambda =\wb\Lambda = \sqrt{\epsilon\eta}\, .
\ee
In this vacuum the $U(1)$ gauge symmetry is broken, and all fields are
massive. Thus the electric flux along $y^1$ is screened, and we get a
supersymmetric ground state. Hence we have a unique supersymmetric ground
state, showing that the number $K$ is indeed 1, as predicted by
$U$-duality.

The $U$-dual of more general $(m,n)$ state described before are given by
bound states of $m$ D- membranes lying in the $8-9$ plane, and $n$
D-membranes lying in the $6-7$ plane. The collective coordinate dynamics
of this system has been described in ref.\cite{BERVAFA}, and the
interacting part of the theory is now given by an $N=4$ supersymmetric
$U(1)\times SU(m)\times SU(n)$ theory in two dimensions, with a
hypermultiplet each in the adjoint representations of $SU(m)$ and $SU(n)$,
and a charged hypermultiplet that carries $U(1)$ charge, and belongs to
the fundamental representations of both, $SU(m)$ and $SU(n)$. It will be
extremely interesting to count the number of supersymmetric ground states
of this system and see if this agrees with the corresponding number
$d(mn)/(16)^3$, which is the number of short multiplets in the elementary
string spectrum for $N_L=mn$.

I would like to thank C. Vafa for useful correspondence.


\begin{thebibliography}{99}

\bibitem{HULL}
C. Hull and P. Townsend, Nucl. Phys. {\bf B438} (1995) 109
(hep-th/9410167).

\bibitem{POLNEW}
J. Polchinski, preprint hep-th/9510017.

\bibitem{WITTENNEW}
E. Witten, preprint hep-th/9510135.

\bibitem{LI}
M. Li, preprint hep-th/9510161.

\bibitem{DBRANE}
A. Sen, preprint hep-th/9510229.

\bibitem{BERVAFA}
M. Bershadsky, V. Sadov and C. Vafa, preprint hep-th/9510225.

\bibitem{SENVAFA}
A. Sen and C. Vafa, preprint hep-th/9508064.

\bibitem{VAFAP}
C. Vafa, private communication.


\end{thebibliography}
\end{document}